\documentclass[]{tPHM2e}

\newcommand{\rme}{{\rm e}}
\newcommand{\rmi}{{\rm i}}
\newcommand{\rmd}{{\rm d}}

\newcommand{\pprime}{{\prime\prime}}

\newcommand{\bra}{\langle}
\newcommand{\ket}{\rangle}

\newcommand{\order}{{\mathcal O}}

\newcommand{\bnull}{{\mbox{\boldmath $0$}}}
\newcommand{\be}{\begin{equation}}
\newcommand{\ee}{\end{equation}}
\newcommand{\bd}{\begin{displaymath}}
\newcommand{\ed}{\end{displaymath}}
\newcommand{\vsp}{\vspace*{3mm}}

\newcommand{\bh}{\ensuremath{\mathbf{h}}}

\newcommand{\bu}{\ensuremath{\mathbf{u}}}

\newcommand{\bx}{\ensuremath{\mathbf{x}}}
\newcommand{\by}{\ensuremath{\mathbf{y}}}
\newcommand{\bz}{\ensuremath{\mathbf{z}}}

\newcommand{\bS}{\ensuremath{\mathbf{S}}}

\newcommand{\bpsi}{{\mbox{\boldmath $\psi$}}}
\newcommand{\bphi}{{\mbox{\boldmath $\phi$}}}

\newcommand{\bsigma}{{\mbox{\boldmath $\sigma$}}}
\newcommand{\btheta}{{\mbox{\boldmath $\theta$}}}

\begin{document}
\doi{}
\issn{1478-6443}
\issnp{1478-6435}
\jvol{00} \jnum{00} \jyear{2011} %\jmonth{21 December}

\markboth{A.C.C. Coolen and K. Takeda}{Philosophical Magazine}

\articletype{}

\title{Transfer operator analysis of the parallel dynamics of \\disordered Ising chains}

\author{Anthony C.C. Coolen$^{\rm a,b}$$^{\ast}$\thanks{$^\ast$Corresponding author. Email: ton.coolen@kcl.ac.uk\vspace{6pt}} 
and Koujin Takeda$^{\rm c}$\\
\vspace{6pt}  
$^{\rm a}${\em{Department of Mathematics, King's College London, The Strand, London WC2R 2LS, UK}}; 
$^{\rm b}${\em{London Institute for Mathematical Sciences, 22 South Audley St, Mayfair, London W1K 2NY, UK}};
$^{\rm c}${\em{Department of Computational Intelligence and Systems Science,  Tokyo Institute of Technology,
 4259 Nagatsuda, Midori-ku, Yokohama 226-8502, Japan}}
}

\maketitle

\begin{abstract}
We study the synchronous stochastic dynamics of the random field and random bond Ising chain. For this model the generating
functional analysis methods of De Dominicis leads to a formalism with transfer operators, similar to transfer matrices in equilibrium studies,
but with dynamical paths of spins and (conjugate) fields as arguments, as opposed to replicated spins. In the thermodynamic limit the macroscopic dynamics is captured by the dominant eigenspace of the transfer operator, leading to a relative simple and transparent set of equations that are 
easy to solve numerically. Our results are supported excellently by numerical simulations. 
\bigskip

\bigskip

\end{abstract}

\section{Introduction}

In spite of the absence of equilibrium phase transitions at finite temperature in one-dimensional Ising chains, 
the dynamics of such systems (solved formally several decades ago \cite{Glauber,Felderhof}) continue to  
be of interest in the context of ageing phenomena, see e.g. \cite{GodrecheLuck}. Disordered versions of such chains, 
with random bonds and/or random fields, generally require 
new techniques for solution,  unless the disorder can be transformed away as for binary bonds.  
One method for solving disordered chains in equilibrium is based on iteration of partition functions for growing chains, 
constrained by the state of the last spin   
\cite{rujan,bruinsma,mukamel,Derrida,Skantzos}. More recently such models were also solved by diagonalisation of replicated transfer matrices
\cite{Nikoletopoulos1,Nikoletopoulos2}. The situation with the  dynamics of disordered Ising chains is less satisfactory.
Except for special cases, e.g. \cite{Forcags}, our analytical methods are still under development, although it is clear from numerical simulations 
(e.g. \cite{Rieger,Corberi}) and from the equilibrium solution that the dynamical phenomenology of disordered Ising chains is rich. 
A renormalisation group approach was advocated in \cite{Fisher_etal}. 
In this paper we use the generating functional method of \cite{Dedom78} to handle the disorder, and show that this leads to a transfer operator formalism very similar to that found in equilibrium studies; we use parallel dynamics to keep computations simpler, but results for Glauber dynamics will be similar.  While developing our study, another study was published \cite{NeriBolle}, also with parallel dynamics but based on the cavity method, 
which appears to represent an alternative but mathematically equivalent perspective on some of our equations.

\clearpage
\section{Model definitions}

We consider $N$ Ising spins $\sigma_i \in \{-1,1\}$ on a periodic one-dimensional chain.
Their dynamics are given by a synchronous stochastic alignment to
local fields of the form $h_i(\bsigma;t)=J_i \sigma_{i-1}+J_{i+1}\sigma_{i+1}+\theta_i(t)$, with the convention $i+N\equiv i$ for all $i\in\{1,\ldots,N\}$, and with
$\bsigma=(\sigma_1,\ldots,\sigma_N)$. Upon defining
$p_t(\bsigma)$ as the probability to find the system at time $t$
in state $\bsigma$, this Markovian process can be written as
\be
p_{t+1}(\bsigma)=\sum_{\bsigma^\prime}
W_t[\bsigma;\bsigma^\prime]p_t(\bsigma^\prime),~~~~~~~~
W_t[\bsigma;\bsigma^\prime]=\prod_i\frac{e^{\beta \sigma_i
h_i(\bsigma^\prime;t)}}{2\cosh[\beta h_i(\bsigma^\prime;t)]}.
 \label{eq:dynamics}
  \ee
 The parameter $\beta=T^{-1}\geq 0$ measures the noise in the dynamics, which 
is fully random for $\beta=0$ and fully deterministic for
 $\beta\to\infty$.
 The $\theta_i(t)$
represent external fields of the form $\theta_i(t)=\theta_i+\tilde{\theta}_i(t)$, with random frozen parts $\theta_i$ and weak time
dependent perturbations $\tilde{\theta}_i(t)$ that serve to define response functions.
The bonds $J_{i}$ and the frozen fields $\theta_i$ are regarded as quenched disorder, drawn for each site $i$ independently from
a distribution $\tilde{P}(J,\theta)$.
 We write averages over the process
(\ref{eq:dynamics}) as $\bra \cdots\ket$ and averages over the
disorder as $\overline{\cdots}$.
Upon removing the time dependent parts of the external fields, so that 
 $h_i(\bsigma;t)=J_i \sigma_{i-1}+J_{i+1}\sigma_{i+1}+\theta_i$,
the process (\ref{eq:dynamics})
obeys detailed balance, and the equilibrium state will be of the Peretto \cite{Peretto} form
\begin{eqnarray}
p(\bsigma)&=& Z^{-1}\rme^{-\beta \tilde{H}_\beta(\bsigma)},
\\
\tilde{H}_\beta(\bsigma)&=&-\sum_{i}\theta_i\sigma_i-\frac{1}{\beta}\sum_{i}\log 2 \cosh[\beta
h_i(\bsigma)].
\end{eqnarray}
The correlation and response functions
$C_{ij}(t,t^{\prime})=\bra \sigma_i(t)\sigma_j(t^{\prime})\ket$ and
$G_{ij}(t,t^{\prime})=\partial\bra\sigma_i(t)\ket/\partial\tilde{\theta}_j(t^{\prime})$ will be related by the 
FDT (fluctuation-dissipation theorem) \cite{review}
\be
G_{ij}(\tau>0)= -\beta[C_{ij}(\tau+
1)-C_{ij}(\tau-1)],~~~~~~G_{ij}(\tau\leq 0)=0.
\label{eq:FDT}
\ee

\section{Generating functional analysis}

In order to analyse the macroscopic dynamics of the chain we concentrate on the calculation of the
disorder averaged generating functional proposed in \cite{Dedom78}: 
\begin{eqnarray}
\overline{Z[\bpsi]} &=& \overline{\bra \exp[-i\sum_i \sum_{t<
t_m} \psi_i(t) \sigma_i(t)] \ket} \nonumber\\ &=&
\sum_{\bsigma(0)} \ldots \sum_{\bsigma(t_m)}
\overline{P[\bsigma(0),\ldots,\bsigma(t_m)] \exp[-i\sum_i \sum_{t<t_m}
\psi_i(t) \sigma_i(t)]}.
\label{eq:Z}
\end{eqnarray}
We isolate the local fields at times $t\in\{0,\ldots, t_m-1\}$ in the usual manner  via  delta
functions, using the short-hand  $\{ \rmd\bh \rmd\hat{\bh}\} = \prod_{i}\prod_{t<t_m}[\rmd h_i(t)
\rmd\hat{h}_i(t)/2\pi]$, which gives 
\begin{eqnarray}
\overline{Z[\bpsi]} &=& \int \{ \rmd\bh \rmd\hat{\bh}\}\sum_{\bsigma(0)}
\ldots \sum_{\bsigma(t_m)} p(\bsigma(0))
e^{N\mathcal{F}[\{\bsigma\}, \{ \hat{\bh}\}]} \nonumber \\
&&\times \prod_{i}\prod_{t<t_m}\rme^{\rmi \hat{h}_i(t)[h_i(t)-\tilde{\theta}_i(t)] -
i\psi_i(t) \sigma_i(t) + \beta \sigma_i(t+1) h_i(t) - \log 2
\cosh[\beta h_i(t)]},
 \label{eq:Zwithfields}
\end{eqnarray}
with the disorder dependent exponent
\begin{eqnarray}
\mathcal{F}[\{\bsigma\}, \{ \hat{\bh}\}] &=& 
\frac{1}{N}\log \prod_i\! \int\!\rmd J\rmd \theta~\tilde{P}(J,\theta)
\rme^{-\rmi\sum_{t}\big\{\theta \hat{h}_i(t)+ J[\hat{h}_i(t)\sigma_{i-1}(t)+\hat{h}_{i-1}(t)\sigma_i(t)]\big\}}.
~~
 \end{eqnarray}
To benefit from the linear nature of the chain we write (\ref{eq:Zwithfields}) 
 in terms of the single-site objects
$\bsigma_i=(\sigma_i(0),\ldots,\sigma_i(t_m\!-\!1))$,
$\bh_i=(h_i(0),\ldots,h_i(t_{m}\!-\!1))$, and $\hat{\bh}_i=(\hat{\bh}_i(0),\ldots,\hat{\bh}_i(t_m\!-\!1))$, with analogous definitions for $\tilde{\btheta}_i$ and $\bpsi_i$.  We also introduce the time shift matrix $\bS$, with entries $S_{tt^\prime}=\delta_{t,t^\prime+1}$, and the vector $\bu=(1,\ldots,1)$, so that $\bx\cdot\bS\by=\sum_{t}x(t+1)y(t)$ and $\bu\cdot\bx=\sum_t x(t)$.
For factorised and homogeneous initial conditions $p_0(\bsigma(0))=\prod_i p(\sigma_i(0))$ we then obtain
\begin{eqnarray}
\overline{Z[\bpsi]} &=& \int\!\prod_i [\rmd \bh_i \rmd\hat{\bh}_i]
 \sum_{\bsigma_1\ldots \bsigma_N}\rme^{-\rmi\sum_i[ \hat{\bh}_i\cdot\tilde{\btheta}_i +
\bpsi_i\cdot\bsigma_i] }
\nonumber
\\
&&\times 
\prod_i \bra \bsigma_i,\bh_i,\hat{\bh}_i|M|\bsigma_{i-1},\bh_{i-1},\hat{\bh}_{i-1}]\ket
\label{eq:tm_form},
 \end{eqnarray}
 with a non-symmetric transfer operator $M$, defined via: \vspace*{-2mm}
 \begin{eqnarray}
\bra\bsigma,\bh,\hat{\bh}|M|\bsigma^\prime,\bh^\prime,\hat{\bh}^\prime\ket &=&
 \frac{p(\sigma(0))\rme^{\rmi \hat{\bh}\cdot\bh + \beta [\bsigma\cdot\bS \bh +\sigma(t_m)h(t_m-1)]}}{\prod_{t<t_m}[4\pi\cosh[\beta h(t)]]}
 \nonumber
 \\
 &&\hspace*{0mm} \times
  \int\!\rmd J\rmd\theta~\tilde{P}(J,\theta)
\rme^{-\rmi\theta\bu\cdot \hat{\bh}-\rmi J[\hat{\bh}\cdot\bsigma^\prime+\hat{\bh}^\prime\cdot\bsigma]}.
\label{eq:kernelM}
\end{eqnarray}
Expression  (\ref{eq:tm_form}) is  for $N\to\infty$ dominated by the largest eigenvalue $\lambda_{\rm max}$ of  (\ref{eq:kernelM}), provided its spectrum is discrete at $\lambda_{\rm max}$. In an equilibrium replica analysis \cite{Nikoletopoulos1,Nikoletopoulos2} the relevant kernel would have replicated spins as arguments;
here the arguments are spin `paths', field `paths' and conjugate field `paths' through time.
The fields $\tilde{\btheta}_i$ and $\bpsi_i$ were only introduced for generating perturbations,  so we 
may expand $\overline{Z[\bpsi]}$ in powers of these fields. To do this efficiently we define 
 \begin{eqnarray}
 &&
{\rm Tr}[K] = \int\! \rmd\bh \rmd\hat{\bh} \sum_{\bsigma} 
\bra\bsigma,\bh,\hat{\bh}|K|\bsigma,\bh,\hat{\bh}\ket,  
\\
&&
\bra\bsigma,\bh,\hat{\bh}|S(t)|\bsigma^\prime,\bh^\prime,\hat{\bh}^\prime\ket=\sigma(t)\delta(\bh\!-\!\bh^\prime)\delta(\hat{\bh}\!-\!\hat{\bh}^\prime)
\delta_{\bsigma,\bsigma^\prime},
\\
&&
\bra\bsigma,\bh,\hat{\bh}|\hat{H}(t)|\bsigma^\prime,\bh^\prime,\hat{\bh}^\prime\ket=\hat{h}(t)\delta(\bh\!-\!\bh^\prime)\delta(\hat{\bh}\!-\!\hat{\bh}^\prime)
\delta_{\bsigma,\bsigma^\prime}.
 \end{eqnarray}
 Since $\overline{Z[\bnull]}=1$ for any $\tilde{\btheta}$ and $Z[\bnull]={\rm Tr}[M^N]$ for $\tilde{\btheta}=\bnull$, according to
 (\ref{eq:Z}) and   
 (\ref{eq:tm_form}) respectively,  we can be sure that 
${\rm Tr} [M^N]=1$ and that all terms of order $\tilde{\btheta}$ or order $\tilde{\btheta}^2$ 
in our expansion (which can be written in terms of derivatives with respect to $\tilde{\btheta}$ of $\overline{Z[\bnull]}$) must be zero. 
 Thus we may write our expansion in the form
\begin{eqnarray}
\overline{Z[\bpsi]} &=&1 -\rmi\sum_{it}
\psi_i(t) \frac{{\rm Tr}[S(t)M^N]}{{\rm Tr}[M^N]}
-\frac{1}{2}\sum_{i tt^\prime}\psi_i(t)\psi_i(t^\prime)
\frac{{\rm Tr}[S(t)S(t^\prime)M^N]}{{{\rm Tr}[M^N]}}
\nonumber
\\
&&\hspace*{-0mm}
-\sum_{i<j}\sum_{tt^\prime}\psi_i(t)\psi_j(t^\prime)
\frac{
{\rm Tr}[M^{N+i-j}S(t)M^{j-i}S(t^\prime)]
}
{{\rm Tr}[M^N]}
\nonumber
\\
&&\hspace*{-10mm}
-\sum_{i\neq j}\sum_{tt^\prime}\psi_i(t)\tilde{\theta}_j(t^\prime)
\frac{{\rm Tr}[M^{N-|i-j|} S(t)M^{|i-j|}\hat{H}(t^\prime)]}{{\rm Tr}[M^N]}
+\order(\bpsi^3\!,\tilde{\btheta} \bpsi^2\!,\bpsi\tilde{\btheta}^2).
~~
 \end{eqnarray}
 We may now use the usual  relations  \cite{Dedom78} to express the quantities of interest in the spin chain
 in terms of derivatives of (\ref{eq:Z}), e.g.  $ \overline{\bra \sigma_i(t)\ket}=\rmi\lim_{\bpsi,\tilde{\btheta}\to \bnull}\partial\overline{Z[\bpsi]}/\partial \psi_i(t)$
 \\[-2mm]
 and $ \overline{\bra \sigma_i(t)\sigma_j(t^\prime)\ket}=-\lim_{\bpsi,\tilde{\btheta}\to \bnull}\partial^2\overline{Z[\bpsi]}/\partial \psi_i(t)$,  giving
  \vspace*{-2mm}
 \begin{eqnarray}
m_i(t)=  \overline{\bra \sigma_i(t)\ket}\hspace*{10mm}
&=&~  \frac{{\rm Tr}[S(t)M^N]}{{\rm Tr}[M^N]},
\\
 C_{ij}(t,t^\prime)=\overline{\bra \sigma_i(t)\sigma_j(t^\prime)\ket}&=&~\frac{
{\rm Tr}[M^{N+i-j}S(t)M^{j-i}S(t^\prime)]
}
{{\rm Tr}[M^N]}~~~~(i\leq j),
\\
G_{ij}(t,t^\prime)=\lim_{\tilde{\btheta}\to \bnull} \frac{\partial \overline{ \bra \sigma_i(t)\ket}}{\partial\tilde{\theta}_j(t^\prime)}
&=&~  -\rmi
\frac{{\rm Tr}[M^{N-|i-j|}S(t)M^{|i-j|}\hat{H}^\prime(t^\prime)]}{{\rm Tr}[M^N]}.
 \end{eqnarray}
Left- and right
eigenvectors with different eigenvalues of  
(\ref{eq:kernelM})
are always 
orthogonal, so we can write (\ref{eq:kernelM}) in the form
$M=
\sum_{\lambda}\lambda 
U(\lambda)$, 
in  which the $U(\lambda)$ are eigenspace projection operators\footnote{If the spectrum of $M$ has continuous parts, the eigenvalue sum becomes an integral.}, with $U(\lambda)U(\lambda^\prime)=0$ if $\lambda\neq \lambda^\prime$. 
The operator $M^\ell=\sum_{\lambda}\lambda^\ell
U(\lambda)$ exchanges dynamical information between sites at distance $\ell$.
Hence  $|\lambda|\leq 1$ for all $\lambda$, and since 
${\rm Tr}[M^N]=1$ for any $N$, $M$ must have an eigenvalue $\lambda=1$. 
Provided the largest eigenvalue is isolated in the spectrum, it follows that $\lim_{N\to\infty}M^N=\lim_{N\to\infty}\sum_{\lambda}\lambda^NU(\lambda)=
U(1)$, and the above expressions give \vspace*{-1.5mm}
  \begin{eqnarray}
\lim_{N\to\infty}m_i(t)
&=& {\rm Tr}[S(t)U(1)]/{\rm Tr} [U(1)],
\label{eq:obs1}
\\
\lim_{N\to\infty}
 C_{ij}(t,t^\prime)&=&
{\rm Tr}[U(1)S(t)M^{j-i}S(t^\prime)]/
{\rm Tr}[U(1)]~~~~(i\leq j),
\label{eq:obs2}
\\
\lim_{N\to\infty}
G_{ij}(t,t^\prime)
&=&  -\rmi
{\rm Tr}[U(1)S(t)M^{|i-j|}\hat{H}^\prime(t^\prime)]/{\rm Tr}[U(1)].
\label{eq:obs3}
 \end{eqnarray}
The above quantities are disorder averages of quantities which, by carrying site indices, will not generally be 
 self-averaging. Hence they will not describe the dynamics of an individual realisation of the chain, but averages over many such realisations. 
 In contrast, the following quantities are expected to be self-averaging: \vspace*{-1.5mm}
  \begin{eqnarray}
  m(t)&= 
\lim_{N\to\infty}\frac{1}{N}\sum_i m_i(t)
&= {\rm Tr}[S(t)U(1)]/{\rm Tr} [U(1)],
\label{eq:obs4}
\\
C(t,t^\prime)&=
\lim_{N\to\infty}\frac{1}{N}\sum_i
 C_{ii}(t,t^\prime)&=
{\rm Tr}[U(1)S(t)S(t^\prime)]/
{\rm Tr}[U(1)],
\label{eq:obs5}
\\
G(t,t^\prime)&=
\lim_{N\to\infty}\frac{1}{N}\sum_i 
G_{ii}(t,t^\prime)
&=  -\rmi
{\rm Tr}[U(1)S(t)\hat{H}^\prime(t^\prime)]/{\rm Tr}[U(1)].
\label{eq:obs6}
 \end{eqnarray}\vspace*{-5mm}

 \section{Spectral properties of the transfer operator}
 
  From now on we consider only chains with independently distributed bonds and fields, i.e. $\tilde{P}(J,\theta)=\tilde{P}(J)\tilde{P}(\theta)$.
 This is the natural and technically easier scenario. To study the spectral properties of $M$ it will be helpful to write this operator as
 \vspace*{-2mm}
 \begin{eqnarray}
\bra\bsigma\!,\bh,\hat{\bh}|M|\bsigma^\prime\!\!,\bh^\prime\!,\hat{\bh}^\prime\ket &=&(2\pi)^{\!-t_m}\!P[\bsigma|\bh]
\rme^{\rmi \hat{\bh}\cdot\bh}\!\!
  \int\!\!\rmd J\rmd\theta~\tilde{P}(J,\theta)
\rme^{-\rmi\theta\bu\cdot \hat{\bh}-\rmi J[\hat{\bh}\cdot\bsigma^\prime+\hat{\bh}^\prime\cdot\bsigma]},~~
\label{eq:kernelMalternative}
\end{eqnarray} 
with the probability $P[\bsigma|\bh]$ of a spin exposed to field path $\bh$ to follow path $\bsigma$:
\vspace*{-2mm}
\begin{eqnarray}
P[\bsigma|\bh]&=& p(\sigma(0))\prod_{t<t_m}
\frac{\rme^{\beta \sigma(t+1)h(t)}}{2\cosh[\beta h(t)]}.
\end{eqnarray}

 \subsection{Reduction of left- and right-eigenvectors} 
 
 On the right-eigenvectors $u_{\rm R}$ of (\ref{eq:kernelMalternative}) we carry out the following transformation:
 \vspace*{-1mm}
\begin{eqnarray}
u_{\rm R}(\bsigma,\bh,\hat{\bh})&=& \int\!\frac{\rmd\bx}{\prod_t(2\pi)}
w_{\rm R}(\bsigma,\bh,\bx)\rme^{\rmi\hat{\bh}\cdot\bx}P[\bsigma|\bh].
\end{eqnarray}
Insertion into the eigenvalue equation reveals that
 $w_{\rm R}(\bsigma,\bh,\by)=w_{\rm R}(\bsigma,\bh-\by)$, and after some trivial manipulations we obtain the 
 simplified eigenvalue problem
  \vspace*{-1mm}
\begin{eqnarray}
\lambda
w_{\rm R}(\bsigma,\bh)
&=&
\sum_{\bsigma^\prime}\! \int\! \rmd\bh^\prime w_{\rm R}(\bsigma^\prime\!\!,\bh^\prime)
\! \int\!\!\rmd J\rmd\theta~\tilde{P}(J,\theta)
  \delta[\bh\!-\!\theta\bu\!-\!J\bsigma^\prime]
  P[\bsigma^\prime|\bh^\prime\!\!+\!J\bsigma].~~
  \label{eq:right_eqn}
\end{eqnarray}
 Writing out the  left-eigenvector equation immediately reveals that  $u_{\rm L}(\bsigma,\bh,\hat{\bh})=u_{\rm L}(\bsigma,\hat{\bh})$.
We now carry out a simple Fourier transformation:
 \vspace*{-1mm}
\begin{eqnarray}
u_{\rm L}(\bsigma,\hat{\bh})&=& \int\!\frac{\rmd\bx}{\prod_t(2\pi)}
w_{\rm L}(\bsigma,\bx)\rme^{-\rmi\hat{\bh}\cdot\bx}.
\end{eqnarray}
Insertion into the left-eigenvalue problem then gives
 \vspace*{-1mm}
\begin{eqnarray}
\lambda
w_{\rm L}(\bsigma,\bh)
 &=&\sum_{\bsigma^\prime}\!\int\!\rmd\bh^\prime  w_{\rm L}(\bsigma^\prime\!,\bh^\prime)
 \int\!\!\rmd J\rmd\theta~\tilde{P}(J,\theta) \delta[\bh\!-\!J\bsigma^\prime] P[\bsigma^\prime|\bh^\prime\!+\!\theta\bu\!+\!J\bsigma].
 ~~
  \label{eq:left_eqn}
\end{eqnarray}
The $w_{\rm L,R}(\bsigma,\bh)$ represent distributions of field path contributions, conditioned on 
spin paths $\bsigma$. Given that $\tilde{P}(J,\theta)=\tilde{P}(J)\tilde{P}(\theta)$ they are connected via 
 \vspace*{-2mm}
\begin{eqnarray}
w_{\rm R}(\bsigma,\bh)&=& \int\!\rmd\theta~\tilde{P}(\theta)~ w_{\rm L}(\bsigma,\bh-\theta\bu).
\label{eq:LRconnection}
\end{eqnarray}
To see this we simply define 
the function $w(\bsigma,\bh)=\int\!\rmd\theta~\tilde{P}(\theta)~ w_{\rm L}(\bsigma,\bh-\theta\bu)$ and use   (\ref{eq:left_eqn}) to establish that it 
obeys
 \vspace*{-1mm}
\begin{eqnarray}
\lambda
w(\bsigma,\bh)
 &=&\sum_{\bsigma^\prime}\!\int\!d\bh^\prime  w_{\rm L}(\bsigma^\prime\!,\bh^\prime)
 \! \int\!\!\rmd J\rmd\theta\rmd\theta^\prime \tilde{P}(J,\theta)\tilde{P}(\theta^\prime)
 \delta[\bh\!-\!\theta^\prime\bu\!-\!J\bsigma^\prime] P[\bsigma^\prime|\bh^\prime\!+\!\theta\bu\!+\!J\bsigma]
 \nonumber
\\
 &=&\sum_{\bsigma^\prime}\int\!d\bh^\prime~  w(\bsigma^\prime\!,\bh^\prime)\int\!\rmd J\rmd \theta ~\tilde{P}(J,\theta)
 \delta[\bh\!-\!\theta\bu\!-\!J\bsigma^\prime] P[\bsigma^\prime|\bh^\prime\!+\!J\bsigma].
\end{eqnarray}
Hence $w(\bsigma,\bh)$ obeys (\ref{eq:right_eqn}) and therefore  (\ref{eq:LRconnection}) holds. We are now left with only one eigenvalue problem, 
and upon combining our results we may summarize
 \vspace*{-1mm}
\begin{eqnarray}
u_{\rm L}(\bsigma,\bh,\hat{\bh})&=&\int\!\frac{\rmd\bx~\rme^{-\rmi\hat{\bh}\cdot\bx}
}{\prod_t(2\pi)}
\phi(\bsigma,\bx),
\label{eq:uL}
\\
u_{\rm R}(\bsigma,\bh,\hat{\bh})&=&  P[\bsigma|\bh]\int\!\frac{\rmd\bx~\rme^{\rmi\hat{\bh}\cdot(\bh-\bx)}}{\prod_t(2\pi)}
\int\!\rmd\theta~\tilde{P}(\theta) \phi(\bsigma,\bx\!-\!\theta\bu),
\label{eq:uR}
\end{eqnarray}
\vspace*{-2mm}
with $\phi(\bsigma,\bx)\equiv w_L(\bsigma,\bx)$ to be solved from
\begin{eqnarray}
\lambda
\phi(\bsigma,\bh)
 &=&\sum_{\bsigma^\prime}\int\!\!\rmd\bh^\prime  \phi(\bsigma^\prime\!,\bh^\prime)\!
 \int\!\!\rmd J\rmd\theta~\tilde{P}(J,\theta) \delta[\bh\!-\!J\bsigma^\prime] P[\bsigma^\prime|\bh^\prime\!+\!\theta\bu\!+\!J\bsigma].
 ~~
 \label{eq:ev_phi}
 \end{eqnarray}
  \vspace*{-4mm}
  
  \noindent
 For $\beta\to 0$ one easily calculates that $\phi(\bsigma,\bh)=2^{-t_m}\sum_{\bsigma}\int\!\rmd J \tilde{P}(J) \delta(\bh\!-\!J\bsigma)$ 
 and that the only possible eigenvalue of (\ref{eq:kernelMalternative}) is $\lambda=1$.

\subsection{Physical meaning of the $\lambda=1$ eigenfunctions}

 The fields experienced at site $i$ can be writen as $\bh_i=\bh^{\rm R}_i+J_{i+1}\bsigma_{i+1}+\theta_i\bu$, where $\bh^{\rm R}_i=J_i \bsigma_{i-1}$. Apart from the periodicity constraint, all information communicated to site $i$ from spins at sites $j<i$ is channeled via $\bh^{\rm R}_i$.
The conditional likelihood $P_i(\bh^{\rm R}|\bsigma)$ to observe $\bh^{\rm R}_i=\bh^{\rm R}$ at site $i$, given we know that $\bsigma_i=\bsigma$, thus obeys 
\begin{eqnarray}
P_i(\bh^{\rm R}|\bsigma)&=& \sum_{\bsigma^\prime}\int\! \rmd\bh^\prime~ P_{i-1}(\bh^\prime|\bsigma^\prime)
P[\bsigma^\prime|\bh^\prime+\theta_{i-1}\bu+J_i\bsigma]\delta(\bh^{\rm R}\!-J_i\bsigma^\prime).
\label{eq:cavity}
\end{eqnarray} 
The spin path $\bsigma^\prime$ at site $i\!-\!1$ is prescribed in $P_{i-1}(\bh^\prime|\bsigma^\prime)$, 
so  $P_{i-1}(\bh^\prime|\bsigma^\prime)$ no longer depends on $\theta_{i-1}$ or $J_i$. Hence if we average (\ref{eq:cavity})  over the disorder 
 we obtain  
\begin{eqnarray}
\overline{P_i(\bh^{\rm R}|\bsigma)}&=& \sum_{\bsigma^\prime}\int\! \rmd\bh^\prime~ \overline{P_{i-1}(\bh^\prime|\bsigma^\prime)}
~\overline{P[\bsigma^\prime|\bh^\prime+\theta_{i-1}\bu+J_i\bsigma]\delta(\bh^{\rm R}\!-J_i\bsigma^\prime)}.
\end{eqnarray} 
Disorder averaging removes any site dependence of $\overline{P_i(\bh|\bsigma)}$, hence  $\overline{P_i(\bh|\bsigma)}=\phi(\bh|\bsigma)$, where the latter is now a true conditional probability distribution, although not corresponding to any specific site, giving the final result
\begin{eqnarray}
\phi(\bh|\bsigma)&=& \lim_{N\to\infty} N^{-1}\sum_i \overline{ P_i(\bh|\bsigma)}=\lim_{N\to\infty}N^{-1}\sum_i 
\overline{\bra \delta(\bh\!-\!J_i\bsigma_{i-1})\ket|_{\bsigma_i\!=\!\bsigma}},
\label{eq:meaning_of_phi}
\\[-1mm]
\phi(\bh|\bsigma)&=& \sum_{\bsigma^\prime}\!\int\!\! \rmd\bh^\prime \phi(\bh^\prime|\bsigma^\prime)\!
\int\!\rmd J\rmd\theta~\tilde{P}(J,\theta) \delta(\bh\!-\!J\bsigma^\prime) P[\bsigma^\prime|\bh^\prime\!+\!\theta\bu\!+\!J\bsigma].
\label{eq:ev_again}
\end{eqnarray} 
Equation (\ref{eq:ev_again}) is identical to (\ref{eq:ev_phi}) for $\lambda=1$. Expression (\ref{eq:meaning_of_phi}) obeys causality, i.e. $\phi(\bh|\bphi)$ is independent of $\sigma(t_{\rm max})$. 
It is reasonable to assume that for $\lambda=1$ there is only one solution of (\ref{eq:ev_again}) that obeys causality,  and 
 that non-causal solutions will be ruled out by time boundary conditions. Thus we may  for $\lambda=1$ 
identify $\phi(\bsigma,\bh)=\phi(\bh|\bsigma)$.
Similar arguments underly the cavity approach in \cite{NeriBolle}, from which (\ref{eq:ev_phi}) can be recovered upon
substituting the characteristics of the 1D chain. 

\section{Calculation of observables}

To calculate the observables (\ref{eq:obs4},\ref{eq:obs5},\ref{eq:obs6}) we need the projection operator $U(1)$.  
If we make the reasonable assumption that for $\beta>0$ the $\lambda=1$ eigenspace is not degenerate, we 
may use (\ref{eq:uL},\ref{eq:uR}), ${\rm Tr}[U(1)]=1$, and $\phi(\bsigma,\bh)=\phi(\bh|\bsigma)$ to write
\begin{eqnarray}
\bra \bsigma,\bh,\hat{\bh}|U(1)|\bsigma^\prime\!,\bh^\prime\!,\hat{\bh}^\prime\ket&=& \gamma^{-1}u_R(\bsigma,\bh,\hat{\bh})u_L(\bsigma^\prime\!,\bh^\prime\!,\hat{\bh}^\prime)
\nonumber
\\
&&\hspace*{-25mm}= \frac{1}{\gamma}P[\bsigma|\bh] \int\!\rmd\bx\rmd\bx^\prime
\phi(\bx|\bsigma) 
\phi(\bx^\prime|\bsigma^\prime)\!
\int\!\!\rmd\theta~\tilde{P}(\theta)
\frac{\rme^{\rmi\hat{\bh}\cdot(\bh-\bx-\theta\bu)-\rmi\hat{\bh}^\prime\cdot\bx^\prime}}{(2\pi)^{2t_m}},
\end{eqnarray}
\vspace*{-3mm}

\noindent with\vspace*{-3mm} 
\begin{eqnarray}
\gamma&=& 
\sum_{\bsigma}\int\!\rmd\bh\rmd\hat{\bh}~u_R(\bsigma,\bh,\hat{\bh})u_L(\bsigma,\bh,\hat{\bh})
\nonumber
\\[-1mm]
&=& (2\pi)^{-t_m}
\sum_{\bsigma}\int\!\rmd\theta~\tilde{P}(\theta)\!\int\!\rmd\bx\rmd\bx^\prime~
P[\bsigma|\theta\bu\!+\!\bx\!+\!\bx^\prime]
 \phi(\bx|\bsigma) \phi(\bx^\prime|\bsigma).
 \label{eq:gamma}
\end{eqnarray}
Since $\phi(\bh|\bsigma)$ is independent of $\sigma(t_m)$ and $P[\bsigma|\bh]$ is independent of 
$h(t_m)$, we can sum in (\ref{eq:gamma}) over $\sigma(t_m)$ and integrate over $h(t_m)$ (in that order), resulting in the same expression for $\gamma$ but with the replacement $t_m\to t_m-1$. Further iteration of this process leads to $\gamma= (2\pi)^{-t_m}$. Hence \vspace*{-1mm}
\begin{eqnarray}
\bra \bsigma,\bh,\hat{\bh}|U(1)|\bsigma^\prime\!,\bh^\prime\!,\hat{\bh}^\prime\ket&=& 
P[\bsigma|\bh] \int\!\rmd\bx\rmd\bx^\prime
\phi(\bx|\bsigma) 
\phi(\bx^\prime|\bsigma^\prime)
\nonumber
\\[-1mm]
&&\times (2\pi)^{-t_m}\!
\int\!\!\rmd\theta~\tilde{P}(\theta)
\rme^{\rmi\hat{\bh}\cdot(\bh-\bx-\theta\bu)-\rmi\hat{\bh}^\prime\cdot\bx^\prime}.
\end{eqnarray}
\vspace*{-2mm}

\noindent
We can now write the dynamical observables (\ref{eq:obs4},\ref{eq:obs5},\ref{eq:obs6}) (using integration by parts in the response function, where we take $t>t^\prime$) in the physically transparent form
  \begin{eqnarray}
  m(t)&=& \sum_{\bsigma} \sigma(t) \int\!\rmd\bx\rmd\bx^\prime~\phi(\bx|\bsigma) 
\phi(\bx^\prime|\bsigma)
 \int\!\!\rmd\theta~\tilde{P}(\theta)
  P[\bsigma|\theta\bu\!+\!\bx\!+\!\bx^\prime],
\\
C(t,t^\prime)&=&
\sum_{\bsigma} \sigma(t)\sigma(t^\prime) \int\!\rmd\bx\rmd\bx^\prime~\phi(\bx|\bsigma) 
\phi(\bx^\prime|\bsigma)
 \int\!\!\rmd\theta~\tilde{P}(\theta)
  P[\bsigma|\theta\bu\!+\!\bx\!+\!\bx^\prime],
\\
G(t,t^\prime)&=&  
\sum_{\bsigma}  \sigma(t) \int\!\rmd\bx\rmd\bx^\prime~\phi(\bx|\bsigma) 
\phi(\bx^\prime|\bsigma)\int\!\!\rmd\theta~\tilde{P}(\theta)
\int\!\rmd\bh~\delta(\bh\!-\!\theta\bu\!-\!\bx\!-\!\bx^\prime)
  \frac{\partial P[\bsigma|\bh]}{\partial h(t^\prime)}
  \nonumber
  \\
  &=& \beta \Big\{ C(t,t^\prime\!+\!1)-
  \sum_{\bsigma}  \sigma(t) \int\!\rmd\bx\rmd\bx^\prime~\phi(\bx|\bsigma) 
\phi(\bx^\prime|\bsigma)
\nonumber
\\[-1mm]
&&\hspace*{20mm} \times \int\!\!\rmd\theta~\tilde{P}(\theta)P[\bsigma|\theta\bu\!+\!\bx\!+\!\bx^\prime]
\tanh[\beta (\theta\!+\!x(t^\prime)\!+\!x^\prime(t^\prime))]\Big\}.
 \end{eqnarray}
 \vspace*{-3mm}
 
 \section{Binary bonds and symmetrically distributed random fields}

 Our equations take a simpler  form when the bonds are binary, i.e. for the choice 
 $\tilde{P}(J,\theta)=\tilde{P}(\theta)[\frac{1}{2}(1\!+\!\eta)\delta(J\!-\!1)+\frac{1}{2}(1\!-\!\eta)\delta(J\!+\!1)]$, where $\eta\in[-1,1]$. 
 Insertion into  (\ref{eq:ev_again}) shows that the dynamic order parameter can now be written as
\begin{eqnarray}
\phi(\bh|\bsigma)
&=& \sum_{\bsigma^\prime}\Phi(\bsigma^\prime|\bsigma)\delta(\bh-\bsigma^\prime),
\label{eq:phi_in_Phi}
 \\
\Phi(\bsigma^\prime|\bsigma)&=&\lim_{N\to\infty}N^{-1}\sum_i 
\overline{\bra \delta_{\bsigma^\prime,J_i\bsigma_{i-1}}\ket}|_{\bsigma_i\!=\!\bsigma},
\end{eqnarray}
 \vspace*{-3mm}

\noindent
with  \vspace*{-2mm}
\begin{eqnarray}
\Phi(\bsigma^\prime|\bsigma)&=& \frac{1}{2}(1\!+\!\eta)
 \sum_{\bsigma^\pprime}
\Phi(\bsigma^\pprime|\bsigma^\prime)\int\!\rmd\theta~\tilde{P}(\theta)
 P[\bsigma^\prime|\bsigma^\pprime\!+\!\theta\bu\!+\!\bsigma]
\nonumber
\\[-1mm]
&&
+\frac{1}{2}(1\!-\!\eta)
\sum_{\bsigma^\pprime}
\Phi(-\bsigma^\pprime|-\!\bsigma^\prime)
\int\!\rmd\theta~\tilde{P}(\theta)  P[-\bsigma^\prime|\theta\bu\!-\!\bsigma^\pprime\!-\!\bsigma].
\end{eqnarray}
 If, furthermore, we choose random fields with $\tilde{P}(-\theta)=\tilde{P}(\theta)$ and unbiased initial conditions $p_0(\sigma(0))=\frac{1}{2}$,
  then $ \int\!\rmd\theta\tilde{P}(\theta)P[-\bsigma^\prime|\theta\bu\!-\!\bsigma^\prime\!-\!\bsigma^\pprime]=  \int\!\rmd\theta\tilde{P}(\theta)
  P[\bsigma^\prime|\theta\bu\!+\!\bsigma^\prime\!+\!\bsigma^\pprime]$, and  the operator in  (\ref{eq:ev_again}) of which we need eigenfunctions 
  commutes with the spin-flip operator $(F\Phi)(\bsigma^\prime|\bsigma)=\Phi(-\!\bsigma^\prime|-\!\bsigma)$. 
  We then find that $\Phi(-\bsigma^\prime|-\bsigma)=\Phi(\bsigma^\prime|\bsigma)$, and the relatively simple equation
  \begin{eqnarray}
  \Phi(\bsigma^\prime|\bsigma)&=& \sum_{\bsigma^\pprime}\Phi(\bsigma^\pprime|\bsigma^\prime)  \int\!\rmd\theta ~\tilde{P}(\theta)   P[\bsigma^\prime|\theta\bu+\bsigma+\bsigma^\pprime].
  \label{eq:Phi}
 \end{eqnarray}
    The formulae for the macroscopic observables can now also be simplified. If we use the following identity, which follows directly 
  from (\ref{eq:phi_in_Phi},\ref{eq:Phi}), 
  \begin{eqnarray}
   \int\!\rmd\bx\rmd\bx^\prime~\phi(\bx|\bsigma) 
\phi(\bx^\prime|\bsigma)
 \int\!\!\rmd\theta~\tilde{P}(\theta)
  P[\bsigma|\theta\bu\!+\!\bx\!+\!\bx^\prime]&=& \sum_{\bsigma^\prime}\Phi(\bsigma|\bsigma^\prime)\Phi(\bsigma^\prime|\bsigma),
  \end{eqnarray}
then  we find $m(t)=0$ for all $t$, and 
      \begin{eqnarray}
C(t,t^\prime)&=&
 \sum_{\bsigma} \sigma(t) \sigma(t^\prime)\sum_{\bsigma^\prime}\Phi(\bsigma|\bsigma^\prime)\Phi(\bsigma^\prime|\bsigma),
 \label{eq:simpleC}
 \\
G(t,t^\prime)&=&  
 \beta \Big\{ C(t,t^\prime\!+\!1)-
  \sum_{\bsigma}  \sigma(t)\sum_{\bsigma^\prime\bsigma^\pprime}\Phi(\bsigma^\prime|\bsigma) 
  \Phi(\bsigma^\pprime|\bsigma) 
 \nonumber
\\[-1mm]
&&\hspace*{10mm} \times
 \int\!\!\rmd\theta~\tilde{P}(\theta)
   P[\bsigma|\theta\bu\!+\!\bsigma^\prime\!+\!\bsigma^\pprime]
\tanh[\beta (\theta\!+\!\sigma^\prime(t^\prime)\!+\!\sigma^\pprime(t^\prime))]\Big\}.
 \label{eq:simpleG}
 \end{eqnarray}
 %%%%%%%%%%%%%%%%%%%%%%%%%
  \begin{figure}[t]
\unitlength=0.28mm\hspace*{5mm}
  \begin{picture}(500,175)
  \put(30,0){\includegraphics[width=250\unitlength]{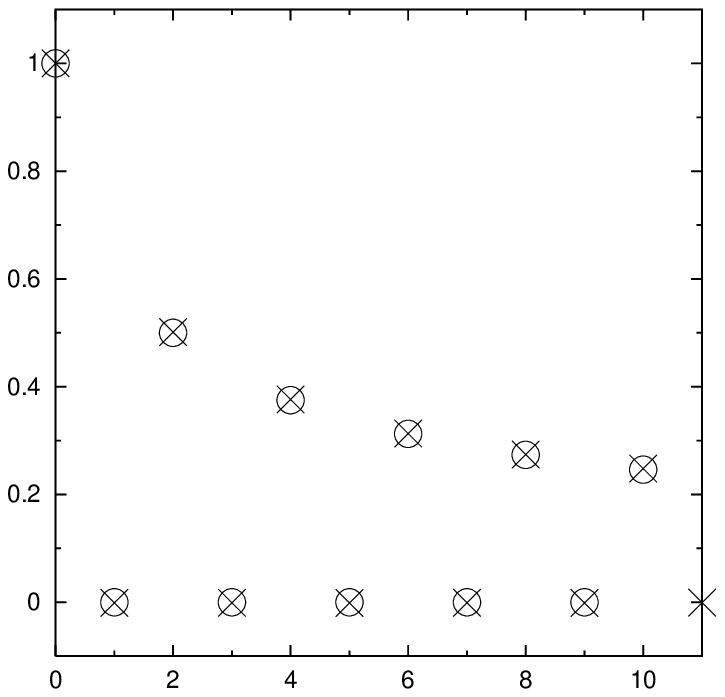}}
  \put(125,-20){$t$} \put(-15,100){$C(0,t)$}
  \put(280,0){\includegraphics[width=250\unitlength]{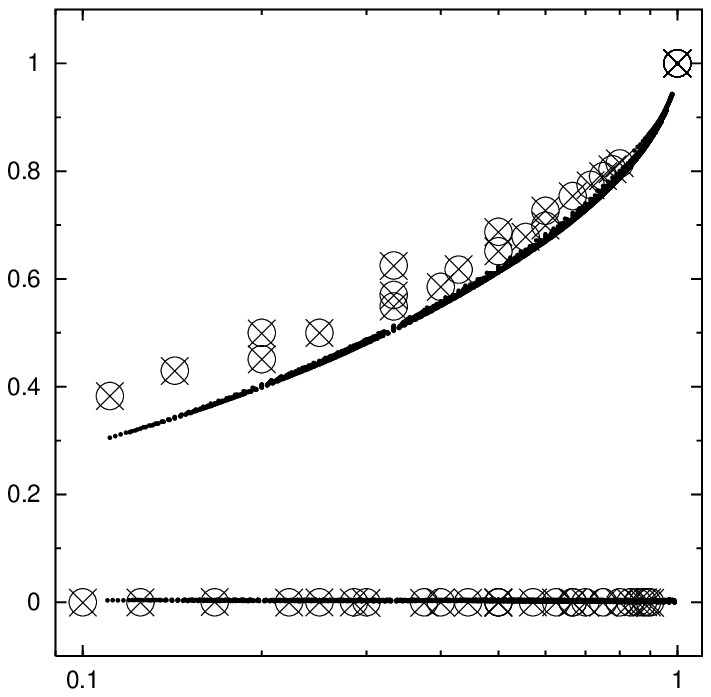}}
  \put(375,-20){$t/t^\prime$} \put(235,100){$C(t,t^\prime)$}
  \end{picture}
  \vspace*{2mm} 
  \caption{Correlations $C(t,t^\prime)$ calculated via numerical solution of  (\ref{eq:Phi}) (for $t,t^\prime\leq 10$, circles) versus 
  correlations measured in numerical simulations with $N=10^6$ (crosses). 
Here  $T=0.1$ and   $\tilde{P}(J,\theta)=\frac{1}{2}\delta(\theta)[\eta \delta(J\!-\!1)+(1\!-\!\eta)\delta(J\!+\!1)]$
(which can be mapped onto the non-disordered Ising chain). 
Left: values of $C(t,0)$, i.e. overlap with the initial state, plotted versus $t$ (which decays as a power law). Right: correlations $C(t,t^\prime)$ plotted versus  the ratio $t/t^\prime$, for $t,t^\prime=1\ldots 10$. 
Dots show the values of $C(t,t^\prime)$ for larger times $50\leq t\leq t^\prime \leq 100$, as measured in simulations, showing the typical nonequilibrium 
behaviour
$C(t,t^\prime)\sim C(t/t^\prime)$ of the ageing regime. Upper branches: even values of $t-t^\prime$; lower branches: odd values of $t-t^\prime$. }
  \label{fig:field00}
  \end{figure}
 \begin{figure} [t]
   \unitlength=0.28mm\hspace*{5mm}
  \begin{picture}(500,175)
  \put(30,0){\includegraphics[width=250\unitlength]{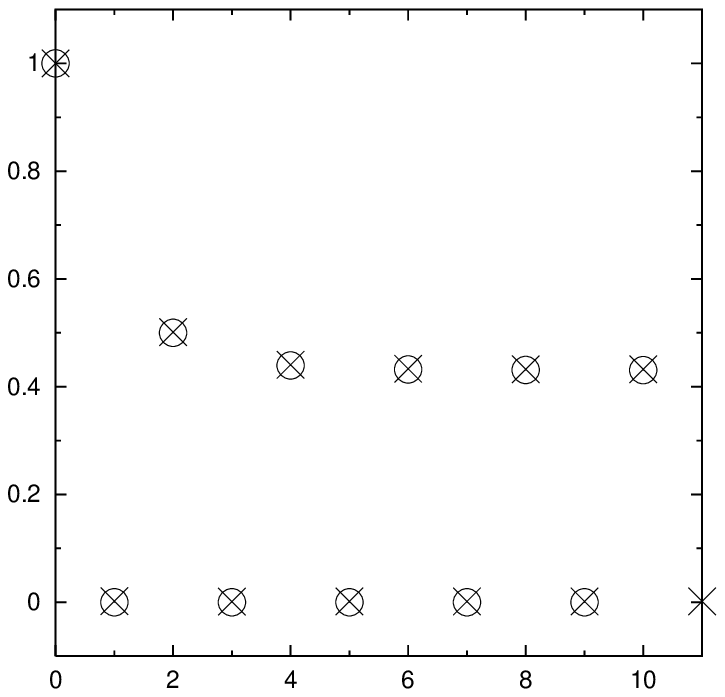}}
  \put(125,-20){$t$} \put(-15,100){$C(0,t)$}
  \put(280,0){\includegraphics[width=250\unitlength]{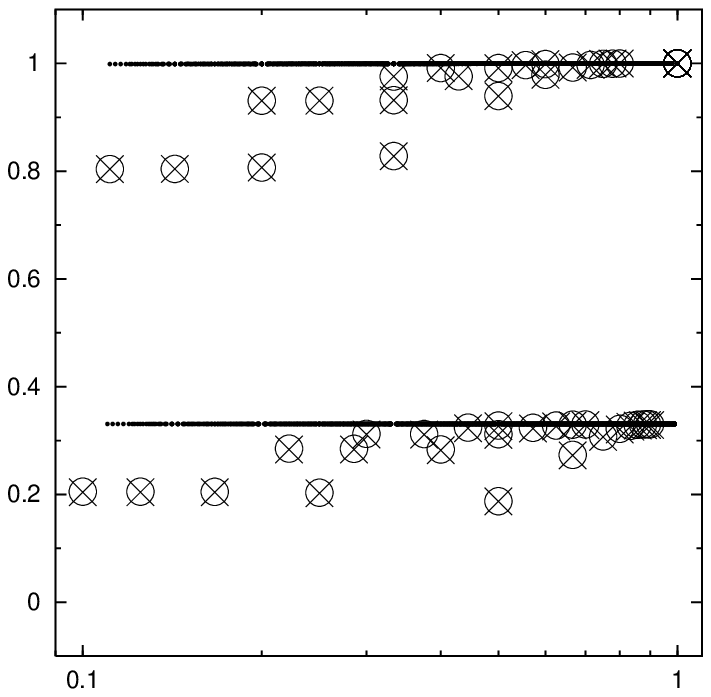}}
  \put(375,-20){$t/t^\prime$} \put(235,100){$C(t,t^\prime)$}
  \end{picture}
  \vspace*{2mm} 
 \caption{Correlations $C(t,t^\prime)$ calculated via numerical solution of  (\ref{eq:Phi}) ($t,t^\prime\leq 10$, circles) versus 
  correlations measured in numerical simulations with $N=10^6$ (crosses). 
Here  $T=0.1$ and   $\tilde{P}(J,\theta)=\frac{1}{4}[\delta(\theta\!-\!\frac{1}{2})+\delta(\theta\!+\!\frac{1}{2})]\delta(\theta)[\eta \delta(J\!-\!1)+(1\!-\!\eta)\delta(J\!+\!1)]$, i.e. weak random fields. 
Left: values of $C(t,0)$, i.e. overlap with the initial state, plotted versus $t$. Right: correlations plotted versus  the ratio $t/t^\prime$, for $t,t^\prime=1\ldots 10$.
Dots show the values of $C(t,t^\prime)$ for larger times $50\leq t\leq t^\prime \leq 100$, as measured in simulations.  Upper branches: even values of $t-t^\prime$; lower branches: odd values of $t-t^\prime$. The system stabilises into a meta-stable state on the timescales considered, from which it takes significantly more time to escape.
 }
  \label{fig:field05}
  \end{figure}
%%%%%%%%%%%%%%%%%%%%%%%%%%%%%%%%%%%%%%
 Equations (\ref{eq:Phi},\ref{eq:simpleC},\ref{eq:simpleG}) no longer depend on $\eta$, since for symmetric field distributions and with $J_i\in\{-1,1\}$ the observables $C$ and $G$ 
 are invariant under gauge transformations of the type $\sigma_i\to \tau_i\sigma_i$, so the bonds can be transformed away. 

For  $\tilde{P}(\theta)=\frac{1}{2}\delta(\theta-\tilde{\theta})+\frac{1}{2}\delta(\theta+\tilde{\theta})$, with $\tilde{\theta}\geq 0$, we are  studying the  synchronous dynamics random field Ising chain for which the statics was solved in \cite{Skantzos}.
 If $\tilde{\theta}>2$ the dynamics close to $T=0$ will be trivial. Each spin freezes into the direction dictated by its external field, and the $T=0$ order parameter reduces 
to    \begin{eqnarray}
  \Phi(\bsigma^\prime|\bsigma)&=&\frac{1}{4}\Big[  \prod_{t=1}^{t_m}\delta_{\sigma^\prime(t),1}
  + \prod_{t=1}^{t_m}\delta_{\sigma^\prime(t),-1}\Big].
 \end{eqnarray}
 For $\tilde{\theta}<2$ the dynamics remains nontrivial. 
Since $P[\bsigma|\bh]$ and $\Phi(\bsigma^\prime|\bsigma)$ obey causality,  numerical solution of equations  
(\ref{eq:Phi},\ref{eq:simpleC},\ref{eq:simpleG}) is for binary fields quite manageable. 
Examples are shown in figures \ref{fig:field00} and \ref{fig:field05}, and compared to data from simulations of chains with $N=10^6$ spins. For those times for which solution of (\ref{eq:Phi},\ref{eq:simpleC},\ref{eq:simpleG})  is feasible, i.e. $t,t^\prime \leq 10$, 
the agreement with simulation data is seen to be perfect. \vspace*{-3mm}

\section{Beyond single-site observables}

Let us finally turn to the calculation of observables that involve multiple sites. 
It is not difficult to transform the eigenvalue problem for $M$ into an equivalent one involving a self-adjoint transfer operator, with the same spectrum. This implies that all eigenvalues $\lambda$ of $M$ must be real. 
We note that, unlike the maximum eigenvalue of (\ref{eq:ev_phi}), the non-leading eigenvalues may well depend on the upper time $t_m$, so we will from now on write $M(t_m)$ and $U(\lambda|t_m)$ instead of $M$ and $U(\lambda)$. 
Since our formulae for observables cannot depend on which choice is made, provided $t_m$ is equal to or exceeding the largest time argument in the observable,  
we can combine (\ref{eq:obs1},\ref{eq:obs2}), the expansion $M(t)=\sum_{\lambda}\lambda U(\lambda|t)$, and the property ${\rm Tr}[U(1|t)]=1$. This 
 allows us to define and work out the two-site correlation function, with $i<j$:
 \begin{eqnarray}
\tilde{C}_{ij}(t,t)&=& \lim_{N\to\infty}\Big\{\overline{\bra \sigma_i(t)\sigma_j(t)\ket}-
\overline{\bra \sigma_i(t)\ket}~\overline{\bra\sigma_j(t)\ket}
\Big\}
\nonumber\\
&=& 
{\rm Tr}[U(1|t)S(t)M^{j-i}(t)S(t)]
-{\rm Tr}[S(t)U(1|t)]~{\rm Tr}[S(t)U(1|t)]
\nonumber\\[-0mm]
&=& \sum_{\lambda(t)\neq 1}\lambda^{j-i}(t)~
{\rm Tr}[U(1|t)S(t)U(\lambda|t)S(t)].
\end{eqnarray}
If not only $\lambda_{\rm max}=1$ but also  second largest eigenvalue $\lambda_2(t)$ is isolated, then 
the evolving correlation length $\xi(t)$  in the chain can be calculated via
\begin{eqnarray}
1/\xi(t)&=&
-\lim_{L\to\infty}\frac{1}{L}\log \Big[N^{-1}\sum_i \tilde{C}_{i,i+L}(t,t)\Big]
\nonumber\\[-1mm]
&=& 
 -\lim_{L\to\infty}\frac{1}{L}\log
\sum_{\lambda(t)\neq 1}\lambda^{L}(t)~
{\rm Tr}[U(1|t)S(t)U(\lambda(t)|t)S(t)]
\nonumber
\\[-1mm]
&=& -\log \lambda_2(t).
\end{eqnarray}
So the present transfer operator picture gives a transparent (although not necessarily trivial) route towards evolving correlation lengths.  
It is not immediately clear whether and how such formulae could be extracted from the cavity formalism 
\cite{NeriBolle}. \vspace*{-5mm}

\section{Discussion}

We have shown that application of the elegant generating functional analysis method of \cite{Dedom78}  to disordered Ising chains (with random fields and/or random bonds) leads to a dynamical version of the familiar 
transfer matrix formalism used in equilibrium studies, with a transfer operator whose arguments are spin paths, field paths and conjugate field paths. Under weak assumptions 
(e.g. isolated largest eigenvalue in the spectrum of the transfer operator) one can take the thermodynamic limit and find an exact self-consistency equation for a dynamical order parameter, from which disorder-averaged single-site correlation and response functions can be calculated explicitly. The latter equation can also be derived from cavity arguments \cite{NeriBolle}, but without the appealing connection with the equilibrium transfer matrix formalism. 
As expected, solving the dynamical order parameter equation is still nontrivial. In this paper we have focused on establishing the principles of the method, and we therefore limited ourselves to numerical solution, for short times only. In a future study we hope to take further steps, and calculate e.g. equilibrium forms for the order parameter (using the parallel dynamics FDT relation) as well as the form of the solution in the ageing regime. We have also limited ourselves here to investigating those properties than can be extracted from the eigenvectors of the  operator that correspond to the largest eigenvalue, which implies calculating single-site objects only. 
However, in analogy with the usual procedure for equilibrium transfer matrices one can also calculate multi-site quantities (such as evolving correlation lengths) from the second largest eigenvalue of the transfer operator. Whether and  how the same could be done within the cavity formalism of 
\cite{NeriBolle} is not obvious. Finally, 
our choice to consider parallel dynamics is not critical for the feasibility of the proposed formalism. In the case of sequential (Glauber) dynamics one will find a very similar structure, but with transfer operators that have continuous time paths rather than discrete time paths as arguments. 
\\[3mm]
{\bf Acknowledgements}\\[1mm]
KT was supported by a Grant-in-Aid (no. 18079006) and Program for
Promoting Internationalisation of University Education from MEXT, Japan.
\\[2mm]
{\em This paper is dedicated to Professor David Sherrington on the occasion of his 70th birthday, 
to thank him for many years of science and friendship.}

\end{document}